\documentclass[preprint,12pt]{elsarticle}

\makeatletter
\def\@author#1{\g@addto@macro\elsauthors{\normalsize%
		\def\baselinestretch{1}%
		\upshape\authorsep#1\unskip\textsuperscript{%
			\ifx\@fnmark\@empty\else\unskip\sep\@fnmark\let\sep=,\fi
			\ifx\@corref\@empty\else\unskip\sep\@corref\let\sep=,\fi
		}%
		\def\authorsep{\unskip,\space}%
		\global\let\@fnmark\@empty
		\global\let\@corref\@empty  
		\global\let\sep\@empty}%
	\@eadauthor={#1}
}
\makeatother

\usepackage{amssymb}

\usepackage{graphicx}
\usepackage{epsfig}
\usepackage{epstopdf}
\usepackage{array}
\usepackage{amsmath}
\usepackage{amssymb}
\usepackage{algorithm}
\usepackage{algpseudocode}
\usepackage{booktabs}
\usepackage{caption}
\usepackage{setspace}

\setcitestyle{square}

\begin{document}

\begin{frontmatter}

\title{A Dictionary Approach to Identifying Transient RFI}
\author{Daniel~Czech\corref{cor1}\fnref{label1}}
\ead{daniel.josef.czech@gmail.com} 
\author{Amit~Mishra\fnref{label1}}
\author{Michael~Inggs\fnref{label1}}
\cortext[cor1]{Corresponding author}

\address[label1]{Department of Electrical Engineering, University of Cape Town, Cape Town, South Africa}

\begin{abstract}
As radio telescopes become more sensitive, the damaging effects of radio frequency interference (RFI) become more apparent. Near radio telescope arrays, RFI sources are often easily removed or replaced; the challenge lies in identifying them. Transient (impulsive) RFI is particularly difficult to identify. We propose a novel dictionary-based approach to transient RFI identification. RFI events are treated as sequences of sub-events, drawn from particular labelled classes. We demonstrate an automated method of extracting and labelling sub-events using a dataset of transient RFI. A dictionary of labels may be used in conjunction with hidden Markov models to identify the sources of RFI events reliably.  We attain improved classification accuracy over traditional approaches such as SVMs or a na\"ive kNN classifier. Finally, we investigate why transient RFI is difficult to classify.  We show that cluster separation in the principal components domain is influenced by the mains supply phase for certain sources. 

\end{abstract}

\begin{keyword}

Transient radio frequency interference classification \sep hidden Markov models \sep kernel principal components analysis \sep density-based clustering \sep transient detection.

\end{keyword}

\end{frontmatter}

\section{Introduction}
\label{intro}

Radio frequency interference (RFI) presents an intensifying problem for radio astronomy. The reasons are two-fold: instruments are becoming significantly more sensitive, while new technologies require more spectrum as they proliferate. In this paper, we aim to classify unintentional, transient RFI. Unintentional RFI is the release of RF energy as an unintended byproduct of the normal operation of some device or system. For example, relays, electric motors and fluorescent lights produce RF signals when switched on or off. It is difficult to identify the sources of this type of interference, since the signals are broadband, impulsive, and often extremely short. Furthermore, there is a risk of RFI being misidentified as real astronomical phenomena. In at least one case a transient RFI source was thought to be a potential astronomical source \cite{pet2015ide}. Certain astronomical observations are particularly vulnerable to impulsive RFI, such as those involving pulsars \cite{mc2003, cordes2004}. In radio quiet reserves, different radio telescopes are often under construction in the close vicinity of others which are already beginning operations. In particular, the infrastructure and equipment attendant to the construction work have the potential to generate unintentional RFI. Therefore, it is important to have the ability to identify the sources of such RFI, so that they may be tracked down and removed or replaced (with shielded equipment, for example). 

There are few attempts to characterise or identify impulsive RFI in the literature. In work by Doran \cite{Doran} for example, three attributes were used: time of day, intensity, and the telescope's pointing direction. Another high-level approach was taken in other work \cite{d2006}, where different vehicles were classified according to their RF emissions when their engines were running. An artificial neural network was used to good effect; however, only five different vehicles were tested. In more recent work \cite{wolfaardt2016}, a number of transient sources found at the MeerKAT radio telescope site were classified using Gaussian Mixture Models and k-Nearest Neighbours (kNN) classifiers with high classification accuracy. However, as the authors note, the recorded transients were limited in length as a result of hardware constraints. In our own prior work \cite{s_0}, we recorded a dataset of 9 typical sources of transient RFI and characterised them in statistical terms. We showed how components analysis techniques (specifically kernel principal components analysis) showed promise as a feature selection step and investigated class separability. We were also able to record for the full length of each RFI event. 

The problem of classifying transient signals is not limited to the radio spectrum. Several important classification tasks deal with acoustic transients, for example in speech processing \cite{rabiner}, sonar target classification \cite{barshan2004} and bio-acoustics \cite{cd2004}. The classification approach that we demonstrate in this paper requires signals to be segmented into individual transients. In one paper investigating the automated classification of cricket songs by species \cite{cd2004}, an algorithm for segmenting audio recordings into individual pulses is presented. The algorithm makes use of a dual-threshold approach, in which the signal energy must exceed two thresholds within a set time period for the onset of a pulse to be detected. Prior to classification, feature vectors are acquired using moving windows, measuring a variety of quantities such as pulse length and pulse frequency, among others. In our prior work \cite{s_0} and in this paper, however, we use kernel principal components analysis (kernel PCA) \cite{scho1997, b_kpca} as part of the feature selection step as all of these quantities are encoded in the resultant principal components. Kernel PCA, a nonlinear extension of PCA, is well established as a feature-selection step in classification problems \cite{kpca_f2004, kpca_f2005, kpca_f2007}.

In this paper, our classification approach deals with sequences of labelled transients. We consider transient RFI signals (or events), which consist of sequences of sub-transients, as analogous to spoken words, which consist of sequences of phonemes. Hidden Markov models (HMMs) \cite{rabiner} have proven highly successful at dealing with this type of classification problem in which sequences are constructed from a lexicon of building blocks. For example, in work by Hu, Lim and Brown \cite{HMM_l0}, HMMs are used for handwriting recognition in this manner - HMMs are trained for individual characters built from a lexicon of sub-character strokes. In another example dealing with sign language recognition, different sub-unit types are used to train HMMs \cite{HMM_l1}. HMMs have also been used in conjunction with a dictionary composed of sub-units to classify whale songs \cite{HMM_l3}.

In our previous paper \cite{s_0}, we showed that a time-domain approach is suitable for classifying the sources of unintentional, impulsive RFI events. In this paper we propose a new approach to the identification of these RFI events, inspired in part by speech processing techniques. 

Our current work contains several novelties. We propose a new way to identify RFI in the time-domain by considering each event as a sequence of transients drawn from a common dictionary. A tailored algorithm for developing such a dictionary is presented, and we illustrate how it may be used as a basis for source identification techniques. We show how events may be separated into individual transients, and how these transients may be labelled using unsupervised density-based clustering methods such as DBSCAN \cite{DBSCAN}. In keeping with our speech-processing inspired approach, we show how hidden Markov models may be used to identify the sources of new RFI events.

As far as we are aware, this is the first time such a dictionary-based approach has been applied to the identification of impulsive RFI in the time-domain, with the exception of our prior conference paper \cite{s_1} in which we reported our initial ideas. This paper extends and completes the two main ideas we proposed in two prior conference papers \cite{s_1, s_2}. Furthermore, we believe this is the first time (aside from our prior conference paper \cite{s_2}) that hidden Markov models have been used to identify RFI events in the time-domain, in an approach loosely inspired by speech processing techniques. We assess the performance of our technique by comparing its accuracy to na\"ive kNN and support vector machine (SVM) classifiers. We use a rigorous method of testing, in which an unseen holdout set is kept completely separate from the training set which is used to train hyperparameters via cross-validation. Lastly, we show that the phase of the AC mains supply at the moment an RFI source is switched can influence clustering behaviour in the principal components domain. We illustrate how this affects cluster separability and hence potential classification accuracy. To our knowledge, this investigation has not been attempted before.

The paper is organised as follows: In Section 2, we describe the dataset used and the limited preprocessing steps taken. We provide a motivation for the dictionary approach in Section 3 and describe how a dictionary of transients may be constructed. Section 3 also includes the algorithms for extracting individual transients from events and subsequently labelling them. In Section 4, we show how a dictionary of transients may be used to identify RFI events by using hidden Markov models, and how various parameters may be optimised. Additionally we report the classification accuracy attained and compare the results to more traditional, direct classification approaches. In Section 5, we show how the mains supply phase influences cluster separation in the principal components domain. Finally, conclusions are drawn in Section 6 and acknowledgments given in Section 7.

\section{Data and Preprocessing}

Two datasets are used in this paper. The first consists of 944 individual time-domain recordings of 9 different sources of transient RFI. These sources are common representative devices that might be found near radio telescopes, especially those operating while others are under construction nearby. The devices are listed in Table~\ref{devices_table}. RFI events were captured using an FPGA-based Reconfigurable Open Architecture Computing Hardware (ROACH) board \citep{ROACH}. Further details on how this dataset was obtained, including a statistical analysis, are given in our previous work \cite{s_0}.

\begin{table}[h]
	\renewcommand{\arraystretch}{1.4}
	\centering
	\caption[]{The sources of RFI used in this paper.}
	\vspace{4mm}
	\begin{tabular}{>{\centering\arraybackslash}m{12mm}m{70mm} >{\centering\arraybackslash}m{20mm}}
		\hline
		\textbf{Class} & \textbf{RFI Source} & \textbf{No. events} \\
		\hline
		1 & Compact fluorescent bulb  & 128 \\
		2 & Power tool & 135 \\
		3 & Small step-down transformer & 142 \\
		4 & Cable & 102 \\
		5 & Mechanical relay (700W resistive load) & 128  \\
		6 & Mechanical relay (without load) & 141 \\
		7 & AC motor ($\approx$ 1 kW) & 63 \\
		8 & Small switching power supply unit & 105 \\
		\hline
	\end{tabular}
	\label{devices_table}
\end{table}

Two of the sources are very similar, each being a different compact fluorescent lamp (CFL). In the clustering approach taken in our prior paper \cite{s_0} they were the least separable of all the sources. We amalgamated these two together to form a single CFL class, reasoning that most of the value lies in detecting that a particular source is a CFL. Knowing the brand is of less importance. Additionally, we limited each full RFI event to 950 000 samples, or around 593 $\mu$s, as virtually all of the activity across events took place within this time. If we had included more samples we would have increased the computational overhead of our analyses unnecessarily, since the additional samples are almost entirely devoid of signal. One further step we took was to normalise event amplitude by class. For each recording $r$ in the set of recordings $R$ of a particular source, $r_{normalised} = \frac{r - \mu}{\sigma}$ where $\mu$ and $\sigma$ are the mean and standard deviation respectively of all $r \in R$.  

A second, smaller set of recordings of three of the sources was collected under lab conditions to test the hypothesis that certain underlying physical characteristics account for variations within clusters of events (dealt with in Section~\ref{mains_effect}). 700 time-domain recordings were taken of the switching signals of three different transient RFI sources. Signal capturing was handled with an Agilent MSO9104A mixed signal oscilloscope and a folded dipole antenna with a center frequency of 146 MHz. To avoid aliasing effects, a low-pass filter was used and the sampling rate was set at 2 GS/s. Each capture was $2\times10^6$ samples long. Further details are given in Section~\ref{mains_effect}.

\section{Dictionary Creation}
\label{dictionary}

\begin{figure}[t]
	\centering
	\includegraphics[trim={3cm 0cm 0 15mm}, width = 15cm]{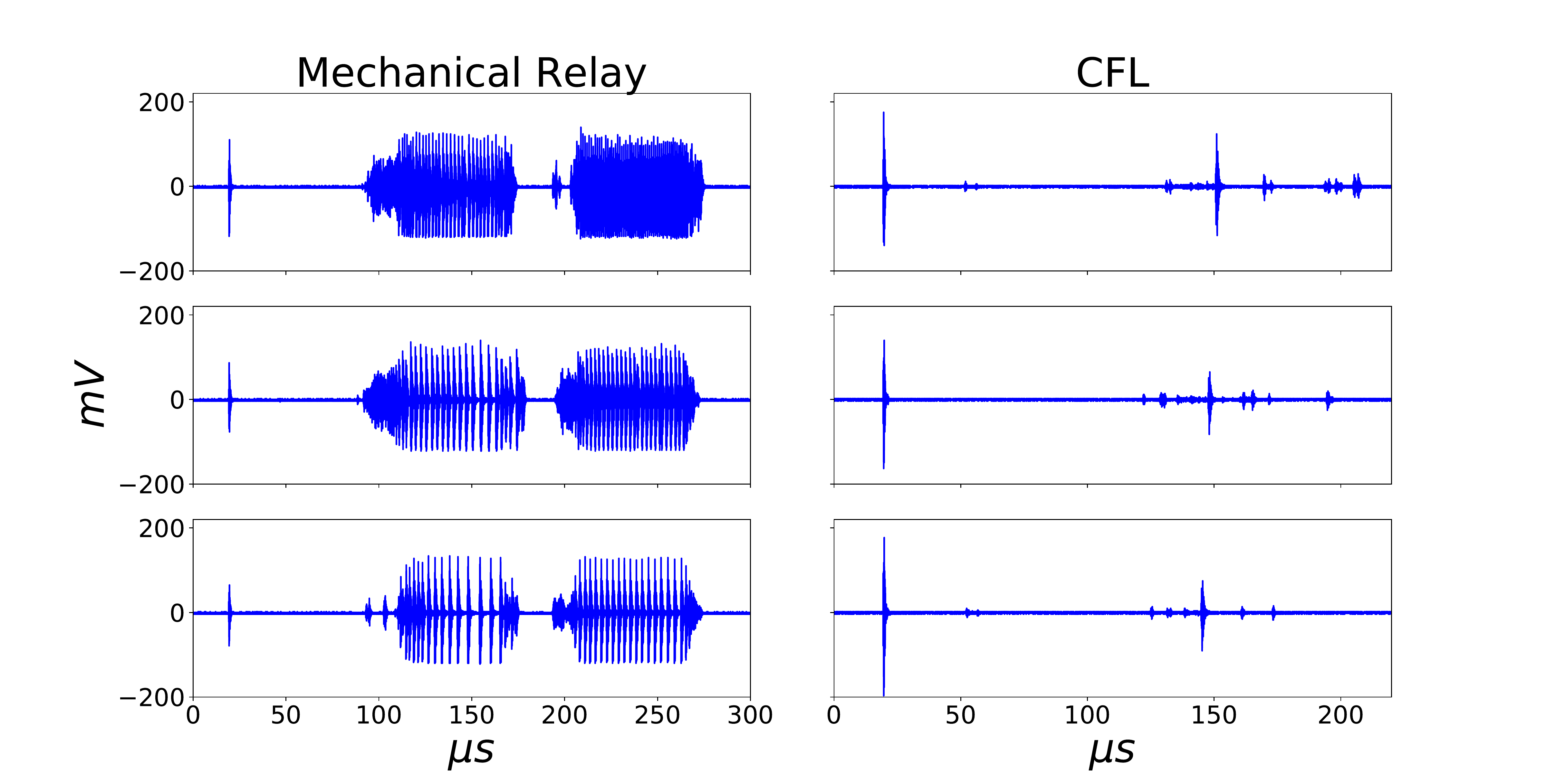}
	\caption{Examples of RFI events caused by switching a mechanical relay (with a resistive load) and a CFL. There is a common underlying structure to the events in each class, although variations exist between them.}
	\label{many_examples}
\end{figure}

In this section, we describe the process of creating a dictionary of transients from which full RFI events may be represented. In our dictionary approach, we propose that full transient RFI events, which consist of sequences of transients, are analogous to spoken words, which consist of sequences of phonemes. We motivate for this approach by making several observations. Firstly, these types of unintentional transient RFI events do tend to consist of sequences of transients. In Fig.~\ref{many_examples} several examples are given. 

Secondly, as with spoken words, there are small variations in repeated events from the same source. Each time a word is spoken, it might sound slightly different, or have a slightly altered sequence of phonemes, but the underlying meaning remains the same. In Fig.~\ref{many_examples}, it is apparent that for a particular RFI source, there are small variations from one recording to the next. However, the underlying structural similarities remain recognisable. Considering RFI events as sequences of transients drawn from a canonical dictionary allows us to use powerful techniques such as hidden Markov models to identify their sources (as we proposed in our prior conference paper \cite{s_2}).  

Furthermore, if this concept of a dictionary is successful, we would expect that RFI events from different sources share at least some of the same labels amongst them. This is the case: in our conference paper \cite{s_1} we showed this for a limited dataset, and in Section~\ref{t_labelling} of this paper, we extend this finding to the full dataset used in this work.    

\subsection{Automated Transient Extraction}
\label{extraction}
As each full event generally contains numerous individual transients, the total quantity extracted from the dataset numbers in the thousands. We deemed it impractical to extract a such a large number of transients by hand, instead developing an automated extraction algorithm and tuning it with a smaller, representative set of hand-selected transients. In this section, we document the approach we developed, although there are surely alternative methods of automatically extracting such transients. We provide the pseudocode in Algorithm~1.

\subsubsection{Extraction Algorithm}
\label{extraction2}

The basic procedure is as follows. Firstly, the Hilbert transform is applied to the raw time-domain recording to extract its approximate envelope. The absolute value of the resultant envelope is smoothed a small amount by convolving it with $L_1$, a moving window. $L_1$ is determined empirically; see Section~\ref{parameter_selection}. A static threshold is applied, set at $F_1 = 1.5\%$ of the maximum range of values in the signal. In this first pass over the signal, a very lenient threshold is permissible, since a stricter threshold is applied later. This minimises the number of transients that are missed while ensuring individual transients are separated from one another. Where this threshold is breached, these regions are selected for further investigation. Regions closer together than $T_M$ samples (determined in Section~\ref{parameter_selection}) are merged and those shorter than $T_M$ samples are discarded. 

\begin{figure}[t!]
	\centering
	\includegraphics[trim={1cm 0cm 0 10mm}, width = 12cm]{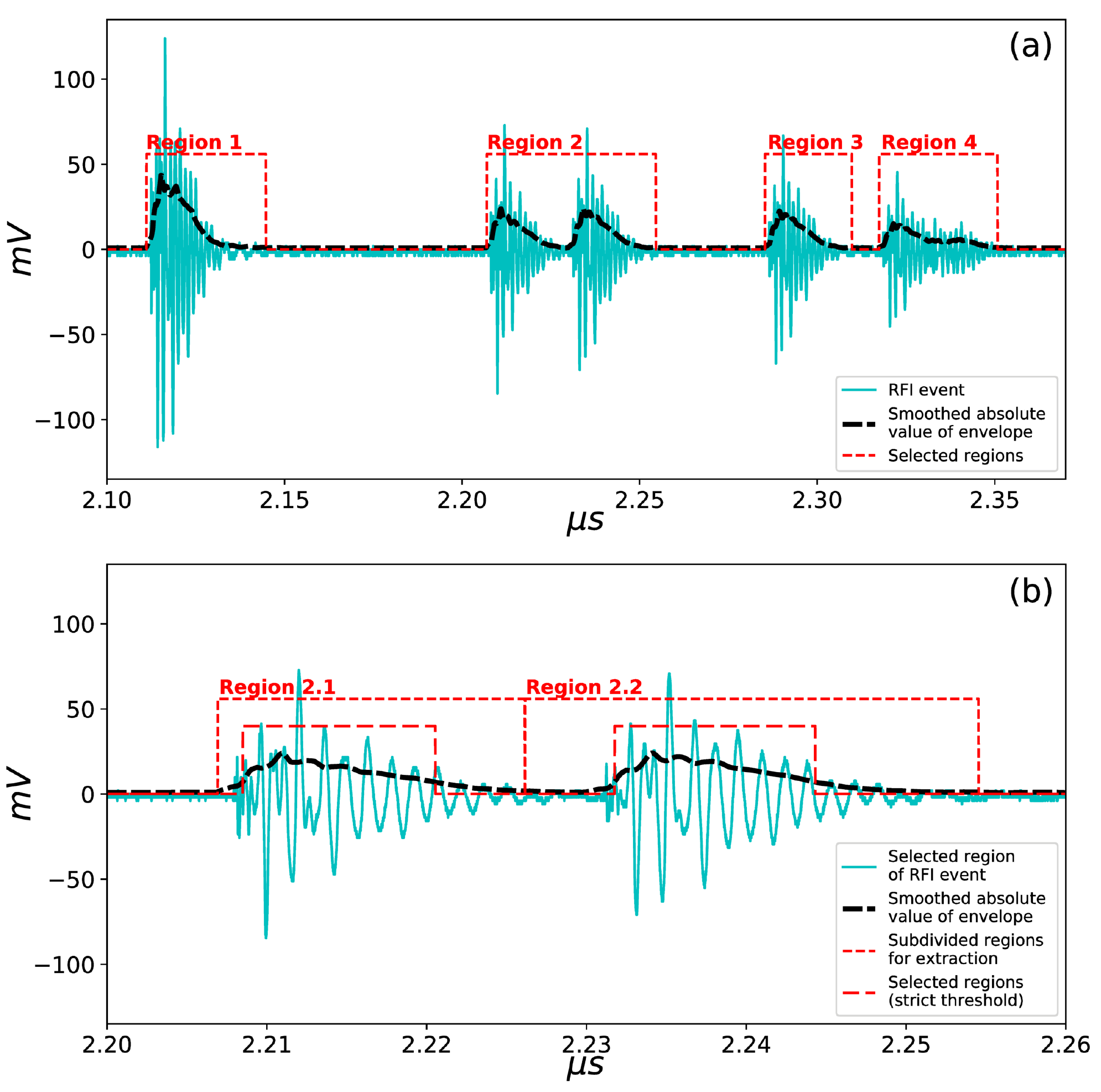}
	\caption{The extraction procedure. A threshold is applied to the smoothed absolute value of the signal's envelope in (a) and (b). In the first pass a lenient threshold is applied, ensuring that all transients are captured. However, such a lenient threshold often results in merged transients, as is the case in Region 2. To avoid this, a second pass with a stricter adaptive threshold is applied to the regions identified in the first pass, as shown in (b). The original regions are then subdivided at the midpoints between subregions resulting from the stricter threshold.}
	\label{RFI}
\end{figure}

For each of the regions selected for further investigation, additional steps are taken to ensure that multiple transients close to one another are not merged together. This procedure is illustrated in Fig.~\ref{RFI}. Firstly, a scaled median filter is applied to the smoothed region extracted previously. The result is an adaptive threshold ($T_2$) that allows us to separate transients that are close together. We used a window length of 8000 samples, although the exact value is not critical, so long as it is kept a few times larger than $L_1$. Subregions are selected where the parent region exceeds this adaptive threshold. In Fig.~\ref{RFI} these subregions are labelled as `Selected regions (strict threshold)'. If there are significant gaps between these subregions, a subdivision is made at the midpoints between them. Finally, segments of the raw time-domain signal (labelled 2.1 and 2.2 in Fig.~\ref{RFI}) are extracted according to these subdivisions.  

	\begin{algorithm}

			\begin{algorithmic}
				\\
				\State \texttt{\textbf{Input:} an RFI event,~} $e(t)$
				\State \texttt{\textbf{Output:} Transients extracted from~} $e(t)$
				\State $e_{H}(t) \leftarrow |H(e_H)(t)|$
				\State $e_{smth}(t) \leftarrow e_{H}(t)*\frac{1}{L_1} \times ones(L_1) $
				\State $e_{mask}(t) \leftarrow $\texttt{where} $e_{smth}(t) > F_1 \times (max(e_{smth}(t))-min(e_{smth}(t))) + min(e_{smth}(t))$
				\For {\texttt{each region} $r_i$ \texttt{in} $e_{mask}(t)$}
				\While{$r_i$ \texttt{closer than} $T_{M}$ \texttt{samples to} $r_{i+1}$}
				\State	$r_i \leftarrow $\texttt{merge} $r_i$ \texttt{and} $r_{i+1}$ 
				\EndWhile
				\If{\texttt{length} $r_i < T_M$}
				\State	\texttt{delete} $r_i$ 
				\EndIf
				\EndFor
				\For {\texttt{each region} $r_i$ \texttt{in} $e_{mask}(t)$}
				\State $e_{smth_i}(t) \leftarrow $ \texttt{region} $r_i$ \texttt{in} $e_{smth}(t)$
				\State $T_2(t) \leftarrow $ \texttt{scaled moving median filter applied to} $e_{smth_i}(t)$
				\State $e_{mask_i}(t) \leftarrow $\texttt{where} $e_{smth_i}(t) > T_2(t)$
				\For {\texttt{each sub-region} $s_j$ \texttt{in} $e_{mask_i}(t)$}
				\State $r_{final} \leftarrow $ \texttt{subdivide} $r_i$ \texttt{at midpoints between} $s_j$ \texttt{and} $s_{j+1}$
				\State  $transient_j(t) \leftarrow r_{final}$ \texttt{in original event signal} $e(t)$
				\State \Return $transient_j(t)$
				\EndFor
				\EndFor 
				
			\end{algorithmic}
	\caption{Transient Extraction }

	\end{algorithm}

\subsubsection{Parameter Selection}
\label{parameter_selection}
The extraction accuracy of the algorithm given in Section~\ref{extraction2} is influenced by several parameters. To optimise them, the accuracy of the extraction algorithm was tested by applying it to a set of transients that were selected by a human. Three metrics were used to assess the performance of the extraction algorithm. 

\begin{enumerate}
	\item The fraction of all ground truth transients that were extracted. A transient is considered extracted if $>60\%$ of its samples (in the manually selected ground truth) are assigned by the extraction algorithm to \textit{any} transient.
	
	\item The fraction of false extractions. An extracted transient is considered falsely extracted if $<30\%$ of its samples belong to a ground truth transient.
	
	\item The fraction of all transients which have been incorrectly merged together by the extraction algorithm.
\end{enumerate}

\begin{figure}[t!]
	\centering
	\includegraphics[trim={0cm 0cm 0 10mm}, width = 13cm]{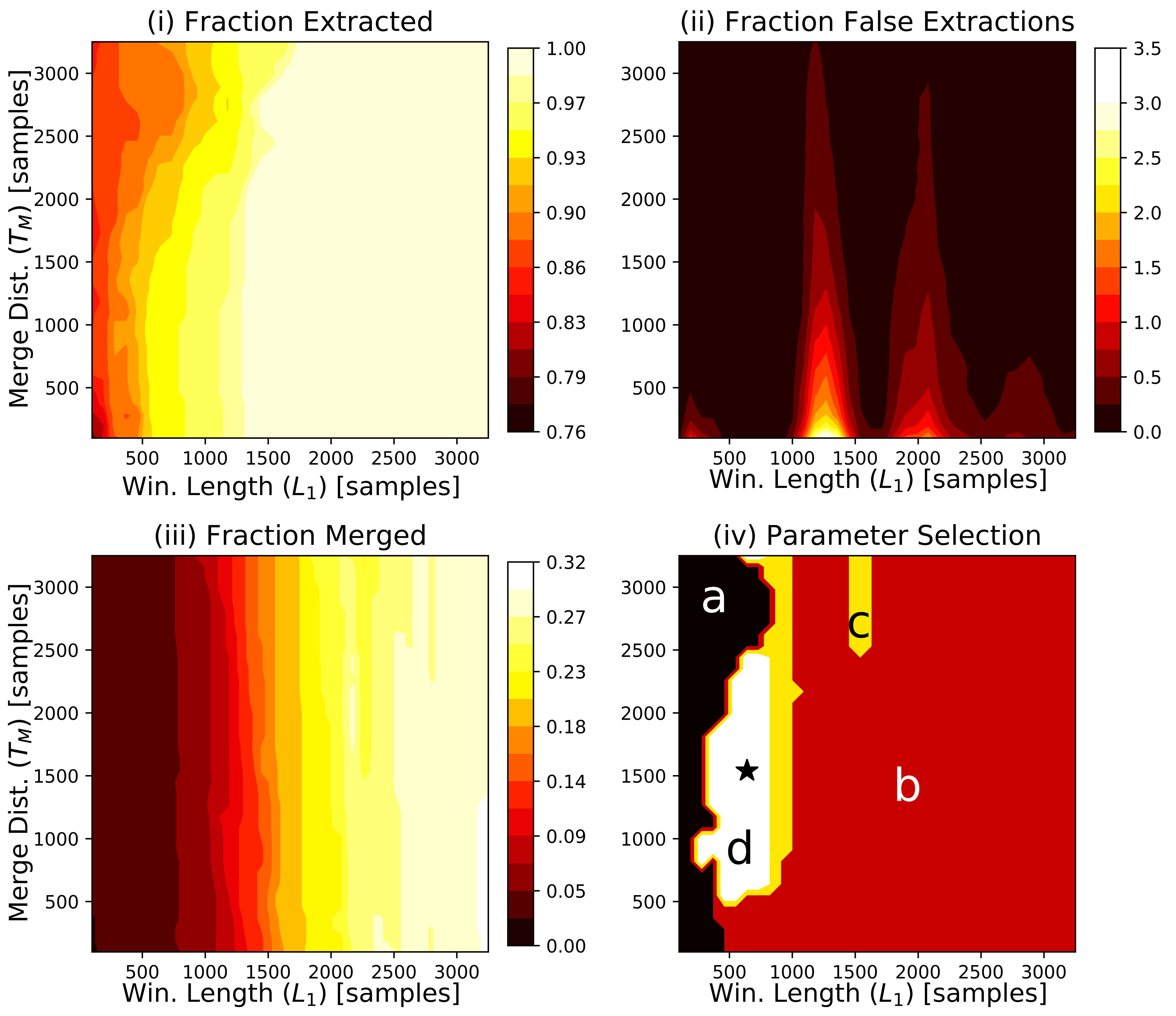}
	\caption{Evaluation results for automated transient extraction. In (i), the fraction of all transients that were correctly extracted for different values of $M_1$ and $L_1$ is given. (ii) shows the number of false extractions as a fraction of the total number of actual transients, a value that can exceed 1. Plot (iii) shows the fraction of all transients that were incorrectly merged together. The final plot indicates how the values of the two parameters were chosen. We set requirements that $>90\%$ of all transients should be extracted, while ensuring that $<5\%$ of all extracted transients are falsely extracted and $<5\%$ of all extracted transients are incorrectly merged. Regions (a), (b) and (c) each fail to satisfy one or more of these criteria, while region (d) satisfies them all. We selected $L_1 = 640$ samples and a $M_1 = 1540$ samples, for an extraction accuracy of $93.4\%$, an erroneous merging rate of $2.6\%$ and a false extraction rate of $2.2\%$. The star indicates where these results are satisfied.}
	\label{optimise_extraction}
\end{figure}

In Fig.~\ref{optimise_extraction}, the extraction results are shown for different values of the two most important parameters, $L_1$ and $T_M$. As can be seen in the plots, there are a wide variety of suitable values of $L_1$ and $T_M$ that offer good results. For the purposes of this investigation, we selected $L_1 = 640$ samples and $T_M = 1540$ samples. Using these values, the extraction accuracy is $93.4\%$, the erroneous merging rate $2.6\%$ and the false extraction rate $2.2\%$.

\subsection{Feature Extraction}

Prior to labelling the transients through unsupervised clustering techniques, we applied a feature selection step to reduce the computational burden. A typical choice might be Principal Components Analysis (PCA). In our previous work \cite{s_0}, we analysed entire RFI events (consisting of multiple transients) and showed that across sample locations, their distributions were non-Gaussian. This led us to conclude that non-linear components analysis techniques such as kernel PCA \cite{scho1997, b_kpca} would fare better than standard PCA (which assumes a Gaussian distribution). Furthermore, we showed that a Gaussian radial basis function is a reasonable choice of kernel function. We can apply these findings to the transients extracted here, since these transients are extracted from the same RFI events studied in our previous paper. Accordingly, we apply kernel PCA to the transients as a feature selection stage, using the following kernel function: 

$$
k(x,y)=e^{-\gamma\|x - y\|^2 }
$$

We set $\gamma = \frac{1}{no.~samples}$ and select the top components (excepting the first 3) that accounted for 65\% of the variance. In addition, prior to unsupervised clustering, we scale each of these components by its relative contribution (within the subset) to the variance in the data. $C$ is the set of $m$ components of dimension $n$, and $\lambda$ is the set of $m$ eigenvalues corresponding to the components in $C$. The components in $C$ are sorted by descending eigenvalue. If $C'$ is the set of scaled components, then for each component $C_i \in C$:

$$
C_i' = C_i\times \frac{\lambda_i - min(\lambda)}{max(\lambda) - min(\lambda)}
$$
\subsection{Transient Labelling}
\label{t_labelling}

Once kernel PCA has been applied, the transients need to be labelled. Of the range of unsupervised clustering methods available, we initially looked at k-means. We quickly realised that density-based methods such as DBSCAN \cite{DBSCAN} would be preferable, however. In the principal components domain, dense, often elongated clusters of points appear within other regions that are relatively sparsely populated with points. As a result the limitations of k-means become quickly apparent - visibly separable clusters are often joined or cut in half depending on the initial conditions of the k-means algorithm.

\begin{figure}[t!]
	\centering
	\includegraphics[trim={5cm 0cm 0 10mm}, width = 14cm]{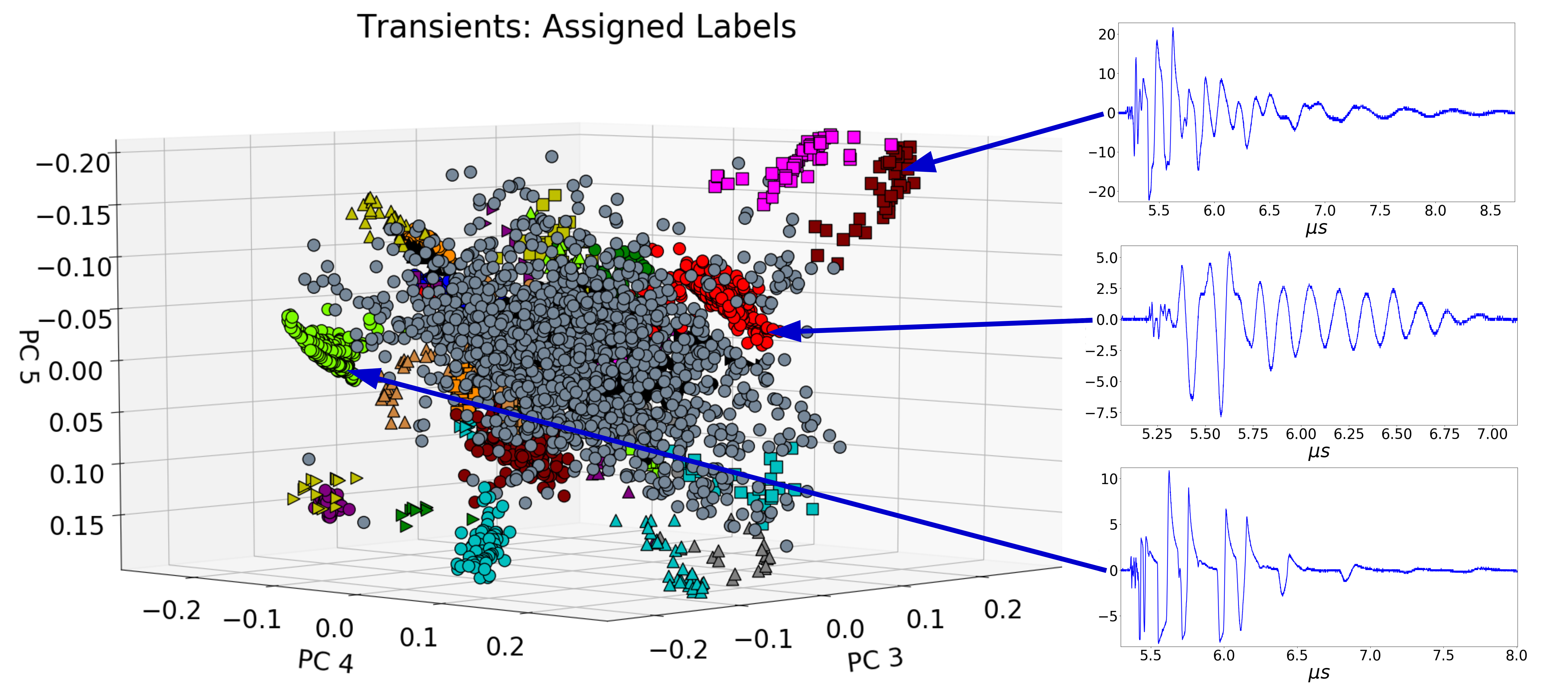}
	\caption{An example of the results of unsupervised clustering to label transients. Each marker represents a particular transient, while the colour and shape of each marker represents the particular label it has been allocated. Note that for legibility we do not show points labelled as noise in this explanatory plot. In addition only 3 particular components are shown here, while the clustering itself was carried out using 92 components. As a result, some clusters which appear to overlap are in fact well-separated in higher dimensions. Three examples of different transient types corresponding to particular labels are given on the right-hand side of the figure. The amplitude of the transients is in arbitrary units, since they are displayed after preprocessing. }
	\label{transient_labels}
\end{figure}

One important disadvantage of DBSCAN is that it struggles to account for variations in cluster density. In the principal components domain, the transients clearly group together in clusters of varying density. As a result, points that belong to sparse (but nevertheless, distinct) clusters are often labelled as noise. There are several variations of DBSCAN that attempt to account for varying cluster density \cite{OPTICS, HDBSCAN} as well as other techniques that use density ratios, for example \cite{drs2016}. We opted to implement a simpler approach in which DBSCAN is performed in a recursive manner. DBSCAN is first applied with parameters tuned for dense clusters. Thereafter it is applied again, this time only on points that were labelled as noise, with parameters favouring sparser clusters. We discuss how we selected these parameters in Section~\ref{performance}. Fig.~\ref{transient_labels} provides a visualisation of our recursive application of DBSCAN to the data, showing 3 of the 92 principal components. We also provide three examples of transients belonging to three of the different clusters.     

If these labels generate a good universal dictionary, we would expect the most important of them to correspond with transients drawn from a wide variety of sources. In other words, we would expect labels to be shared amongst events from different sources. In Table~\ref{distribution}, we show how three of the most prevalent labels (out of 53 in this example) are distributed amongst the 8 classes in this paper. As expected, they account for a large proportion of the transients drawn from each class, and are widely shared amongst classes.

\newcommand{\ra}[1]{\renewcommand{\arraystretch}{#1}}
\begin{table*}[h!]\centering
	\ra{1.3}
	\begin{tabular}{@{} m{3cm} c c c c @{}}\toprule
	
		& \multicolumn{4}{c}{\textbf{ \% of in-class transients}} \\
		\cmidrule{2-5} 
		\textbf{Source} & Label 1 & Label 2 & Label 3 \\ \midrule
		power tool 	& 27.83 & 33.36 & 4.65 	\\
		transformer & 31.26	& 2.68 	& 7.72  \\
		cable 		& 21.66	& 57.49	& 7.77  \\
		relay (load)& 27.48 & 7.62 	& 0.25	\\
		relay 		& 27.69 & 25.82 & 0.0 	\\
		AC motor 	& 27.91	& 42.55 & 1.08 	\\
		CFL 		& 33.11 & 13.33 & 12.80	\\
		PSU 		& 27.79	& 25.81 & 11.79 \\
		\bottomrule \\
	\end{tabular}
	\caption{The distribution of three of the most important labels amongst the 8 classes. A large proportion of the transients within each class correspond with labels widely shared amongst the 8 classes, supporting the concept of a canonical dictionary as proposed here and in our previous conference paper \cite{s_1}.}
	\label{distribution}
\end{table*}

\section{RFI Source Identification}

In this section, we describe how RFI events may be identified using hidden Markov models, based on the labelled dictionary of transients. We also compare the classification accuracy of this dictionary-based approach to two other classic methods, a k-Nearest Neighbours (kNN) classifier and an SVM classifier.

\subsection{Sequence Reconstruction and New Events}
\label{sequence_reconstruction}

\begin{figure}[h!]
	\centering
	\includegraphics[width = 12cm]{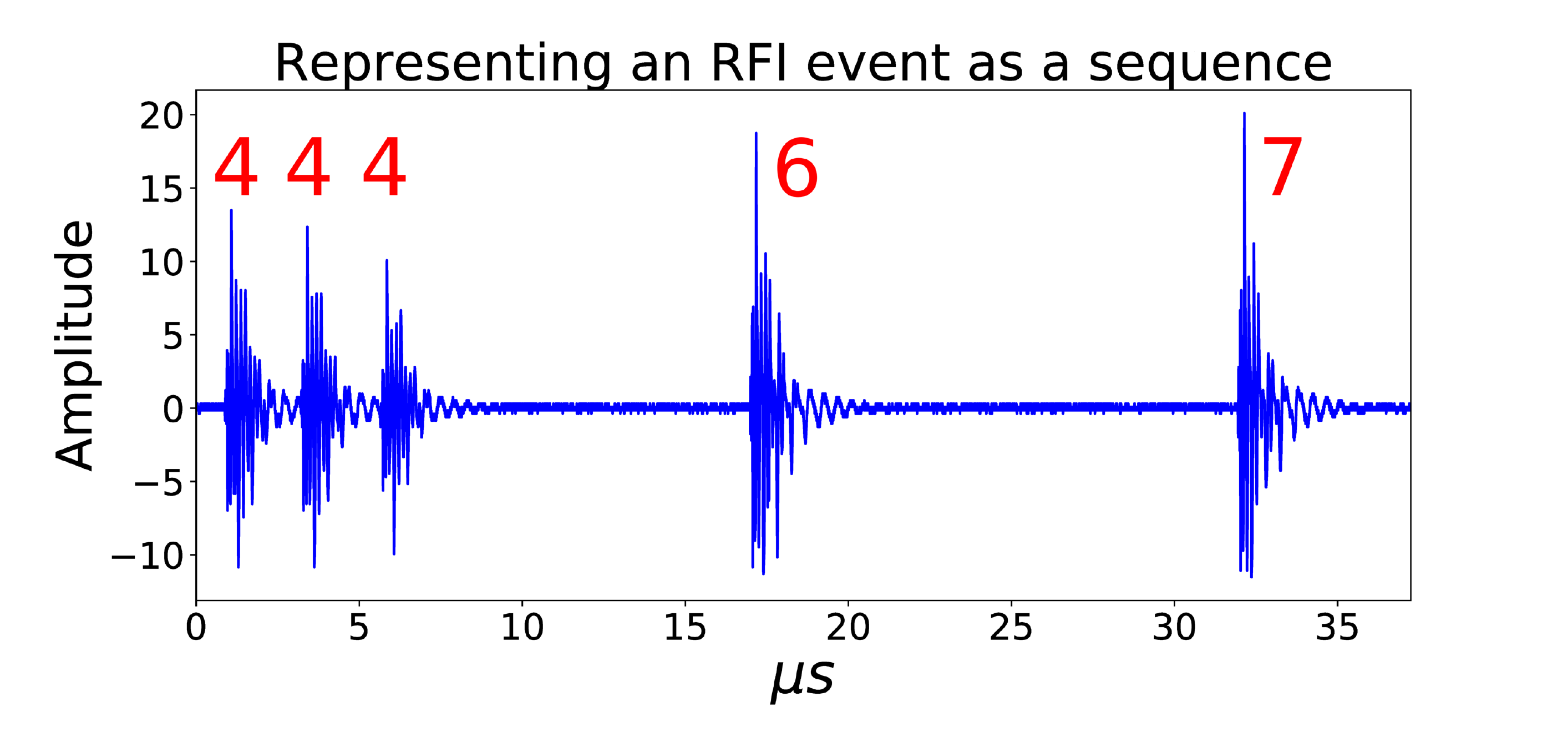}
	\caption{An example of a full RFI event represented as a sequence of labels. Each of the transients is labelled as described in Section~\ref{t_labelling} or~\ref{sequence_reconstruction} depending on whether the RFI event is used in training or testing. The RFI event is then represented as a sequence of these labels: \textbf{[4, 4, 4, 6, 7]} for example. Units of amplitude are not given since this particular example is displayed after preprocessing.}
	\label{seq_rep}
\end{figure}

Now that a dictionary has been constructed, each full RFI event may be represented as a sequence of labelled transients. A full RFI event is given as a sequence of numbers, each number representing the label assigned to a transient in the sequence of transients that together make up the event. An example is given in Fig.~\ref{seq_rep}. New events, those that are not part of a training set from which a dictionary is constructed, are converted to sequences as follows: Firstly, a new raw RFI event is divided into its individual transients using the extraction approach described in Section~\ref{extraction}. These extracted transients are then represented in the principal components space by projecting them onto the principal components extracted previously from the transients in the training data. They are then assigned labels based on the clusters obtained via the pre-constructed dictionary. Specifically, a new point is assigned the label belonging to the most prevalent class in the k-nearest training points. Finally, the new event is represented as a sequence of labels, as with the training data.  

\subsection{Source Identification Using Hidden Markov Models}

Given our approach in which RFI events are analogous to spoken words, and individual transients are analogous to phonemes, HMMs are a natural choice for identification purposes. They are a proven approach to classifying non-stationary and piecewise stationary signals in many fields, such as speech processing \cite{rabiner, HMM2} and financial time-series \cite{HMMF}. Their robustness to variations in repeated examples of the same signal is responsible for much of their success in speech processing. As discussed in Section~\ref{dictionary}, the full RFI events, when considered as sequences of transients, exhibit such variations in repeated examples from the same source. In addition (as noted in Section~\ref{intro}) HMMs are well suited to classification problems involving sequences of sub-units \cite{HMM_l0, HMM_l1}. In our case, the sub-units are labelled transients. We omit the theory of HMMs here for conciseness. For an excellent explanation of the basic theory of HMMs, please consult the paper by Rabiner \cite{rabiner}.  

In Section~\ref{sequence_reconstruction} we described how each full RFI event is represented as a sequence of labels. The sequences reconstructed from events in each randomised training set are used to train an HMM for each different class, using the Expectation-Modification method \cite{rabiner}. When a new event (from a test set) is to be classified, a likelihood figure of merit is calculated from the log probability for each of the different models, using the Viterbi algorithm \cite{rabiner}. The new event is then assigned the class of the model for which the highest likelihood figure of merit is calculated. The optimal number of hidden states, $n$, is determined empirically as a hyperparameter (see Section~\ref{performance}). We used the software package hmmlearn \cite{hmm_learn} to train and apply unconstrained HMMs with Gaussian emission probabilities.

\subsection{Parameter Tuning and Classification Performance}
\label{performance}

Prior to hyperparameter tuning and performance evaluation, the data were divided into 5 randomised sets of approximately 189 examples each. These sets were stratified, in that as near as possible, $\frac{1}{5}$ of the examples from each class were included in each of the 5 sets. One of these sets was then set aside as a hold-out testing set, never to be used in hyperparameter tuning or model training. The remaining four were combined in four different train/development pairs and used in four-fold cross-validation to tune hyperparameters. Each fold was made up of approximately 566 training examples, and 189 validation samples. Due to the large search space, a series of parametric searches were conducted to determine good values for the various hyperparameters. These parameters included the number of neighbouring points, $k$, to use when assigning class labels to new transients (see Section~\ref{sequence_reconstruction}); the number of hidden states to use when training the hidden Markov models, and the two $minpts$ and two $eps$ parameters for each application of DBSCAN. We present the optimal values in Table~\ref{params_table}. Our final accuracy result was obtained by using all four cross-validation sets together as a training set, and testing on the hitherto completely unseen testing set. 

\begin{table}[h]
	\renewcommand{\arraystretch}{1.4}
	\centering
	\caption[]{Classification results and associated parameters. The test accuracy is the classification accuracy on the unseen test-set, while the training accuracy is the average accuracy of four-fold cross-validation applied to the training set.}
	\vspace{3mm}
	\begin{tabular}{m{15mm}m{15mm}m{11mm}m{11mm}m{11mm}m{11mm}m{8mm}m{22mm}}
		\hline
		\% acc. (test) & \% acc. (train) & min- pts1 & min- pts2 & eps1 & eps2 & k & No. hidden states \\
		\hline
		69.47 & 65.54 & 40 & 10 &  0.0745 & 0.0792 & 11 & 33 \\
		\hline
	\end{tabular}
	\label{params_table}
\end{table}

Surprisingly, our unseen test result exceeded the best average score we were able to obtain using four-fold cross-validation on the training set. We attribute this to the $25\%$ increase in training examples when training the final dictionary. During cross-validation, only $60\%$ of the data were available at any one time for training, whereas when training the final dictionary, we could use the full cross-validation set consisting of $80\%$ of the data. In Table~\ref{confusion}, we display a confusion matrix, showing exactly how the test set data were classified. The best result, correctly classified $93.1\%$ of the time, was for the small step-down transformer. The worst was for the AC motor, classified correctly only $38.5\%$ of the time, and misclassified as a power tool and a power supply unit $23.1\%$ of the time each. Misclassification as a power tool is not necessarily bad, however, as the power tool contains a smaller AC motor of a similar type.    

\begin{table}
	\renewcommand{\arraystretch}{2.25}
	\caption[]{ This table displays the classification results for the unseen test set. An overall accuracy of $69.47\%$ was obtained. Each row shows as a percentage how the examples belonging to a particular class were classified. In the perfect case, each of the bold values along the diagonal would be $100\%$, indicating that each source was correctly classified $100\%$ of the time.     
	}
	\vspace{3mm}
	\begin{tabular}{ r|c|c|c|c|c|c|c|c| }
		\multicolumn{1}{r}{}
		&  \multicolumn{1}{c} {\rotatebox[origin=l]{90}{power tool}}
		&  \multicolumn{1}{c}{\rotatebox[origin=l]{90}{transformer}}
		&  \multicolumn{1}{c}{\rotatebox[origin=l]{90}{cable}}
		&  \multicolumn{1}{c}{\rotatebox[origin=l]{90}{relay (load)}}
		& \multicolumn{1}{c}{\rotatebox[origin=l]{90}{relay }} 
		&  \multicolumn{1}{c}{\rotatebox[origin=l]{90}{AC motor}}
		&  \multicolumn{1}{c}{\rotatebox[origin=l]{90}{CFL}}
		&  \multicolumn{1}{c}{\rotatebox[origin=l]{90}{PSU}}\\
		
		\cline{2-9}
		power tool & \textbf{51.9} & 11.1 & 7.4 & 0.0 & 22.2& 3.7 & 0.0 & 3.7\\
		\cline{2-9}
		transformer & 0.0 & \textbf{93.1} & 0.0  & 0.0 & 3.4 & 0.0 & 3.4 & 0.0 \\
		\cline{2-9}
		cable & 0.0 & 0.0 &\textbf{ 80.0} & 5.0 & 10.0 & 5.0 & 0.0 & 0.0\\
		\cline{2-9}
		relay (load) & 0.0 & 0.0 & 3.8 & \textbf{69.2} & 7.7 & 15.4 & 0.0 & 3.8\\
		\cline{2-9}
		relay & 0.0 & 3.6 & 25.0 & 0.0 & \textbf{71.4}& 0.0 & 0.0 & 0.0\\
		\cline{2-9}
		AC motor & 23.1 & 7.7 & 0.0 & 0.0 & 0.0&\textbf{38.5} & 7.7 & 23.1 \\
		\cline{2-9}
		CFL & 0.0 & 0.0 &11.5 & 0.0 & 11.5 & 0.0 &\textbf{73.1}  & 3.8\\
		\cline{2-9}
		PSU & 4.8 & 4.8 & 4.8 & 0.0 & 14.3 & 0.0 & 9.5 &\textbf{61.9} \\
		\cline{2-9}
	\end{tabular}
	\label{confusion}
\end{table}

Another way to evaluate the dictionary-based classification approach is to compare its performance with that of other, well known methods. For the comparison we conduct here, we use the same cross-validation sets and separate unseen holdout set as before. In this case, we do not extract the transients that make up each full event; rather we consider each full event to be one training/development/testing example. Before applying the different classification methods, we use kernel PCA for dimensionality reduction (see our previous work \cite{s_0}), selecting the top components that together explain $80\%$ of the variance in each cross-validation training set. Following this step, we tested two basic methods, SVMs and a kNN classifier. In the case of the kNN classifier, we optimised k using the same cross-validation approach as before. As shown in Table~\ref{comparison_table}, the dictionary-based approach to identifying these RFI events in conjunction with HMMs significantly outperforms these basic methods.   

\begin{table}[h]
	\renewcommand{\arraystretch}{1.4}
	\centering
	\caption[]{Comparison of the dictionary method and other basic methods. A linear SVM (rather than kernel-based SVMs) performed best after kernel PCA had been applied to the data. The best result for the kNN classifier was obtained for $k=1$. }
	\vspace{3mm}

	\begin{tabular}{m{28mm}m{25mm}m{32mm}m{32mm}}
		\hline
		\textbf{Classification method} & Dictionary approach and HMMs & Direct approach and SVM classifier & Direct approach and kNN classifier \\
		\hline
		\textbf{Overall accuracy [\%]} & 69.47 & 58.42 & 52.11\\
		\hline
	\end{tabular}

	\label{comparison_table}
\end{table}

\section{The Difficult Nature of Classifying Transient RFI}
\label{mains_effect}

It is clear that classifying unintentional transient RFI is a difficult task. As we have shown, a traditional approach with SVMs or kNN-based classifiers (along with a kernel PCA feature reduction step) yields poor results. Unravelling the reasons responsible for this will aid the development of future identification algorithms. During our investigations, we began to suspect that the mains supply voltage phase had an effect on cluster shapes in the principal components space. That is, the shape and density of a cluster belonging to a specific device is related to the supply phase at the instant the device is switched. It is already known that the supply voltage affects the generation of impulsive RFI from certain sources - for example, in one paper \cite{2008p} it is shown that impulsive RFI generated by a defective thermostat exhibits a high correlation with the positive and negative peaks of the supply voltage. However, to our knowledge, the effects of supply phase on clusters of events in the principal components domain have not yet been investigated.  

\begin{figure}[t]
	\centering
	\includegraphics[trim={0 3cm 0cm 3cm}, width = \textwidth]{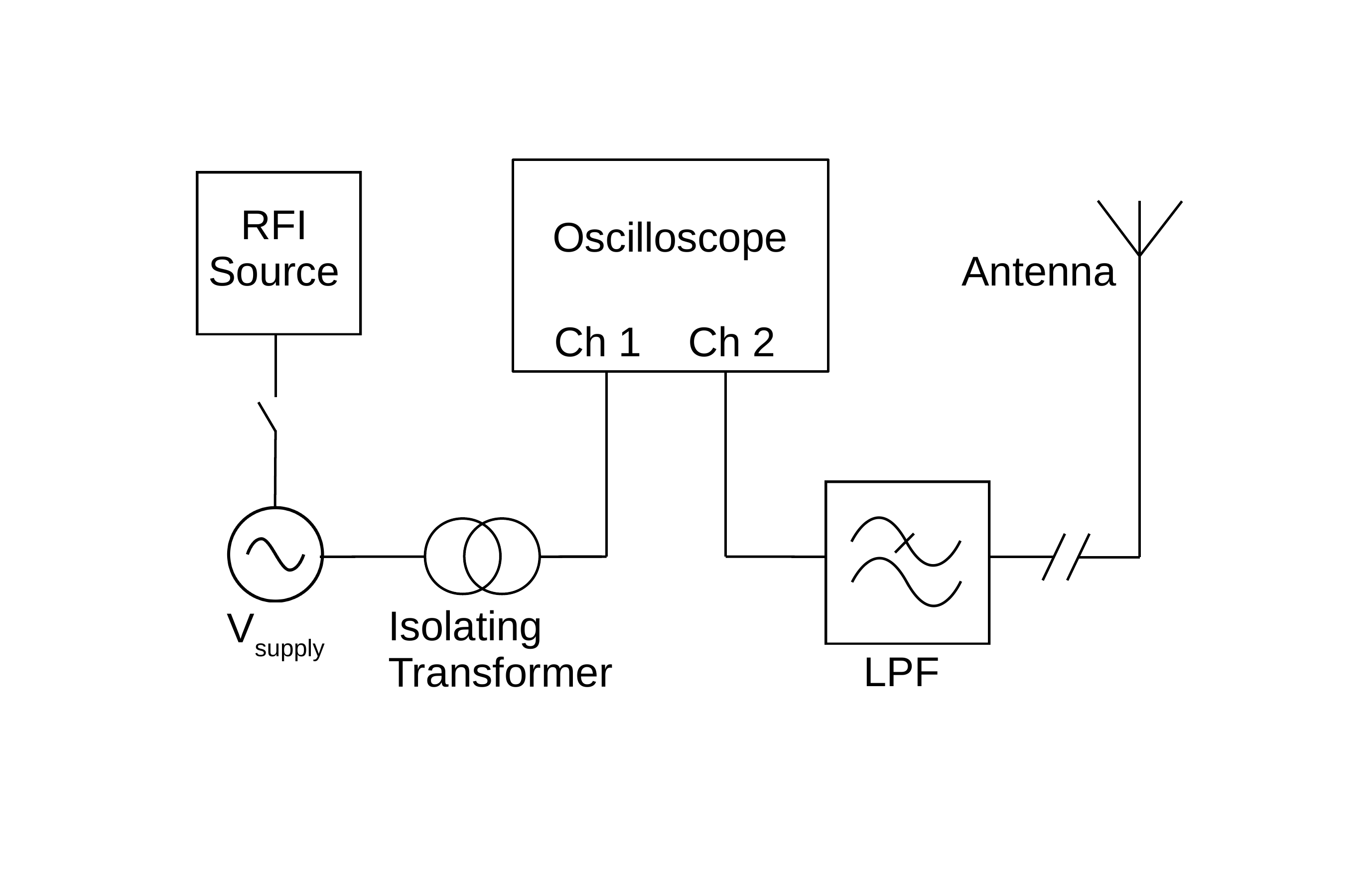}
	\caption{The experiment to record RFI events and the mains supply voltage simultaneously.}
	\label{collection2}
\end{figure}

We set up an experiment in which three of the RFI sources were switched between 200 and 300 times each and the corresponding transient RFI signals recorded. Simultaneously, the instantaneous mains supply voltage was recorded at the moment each device was switched. We used a folded dipole antenna with a center frequency of 146 MHz, identical to the one used in our previous paper \cite{s_0}. Both the RF signal and mains supply measurement were handled with an Agilent MSO9104A mixed signal oscilloscope to ensure simultaneous captures. A low-pass filter was used for anti-aliasing purposes. In Fig.~\ref{collection2}, we provide a diagram of the experiment.

\begin{figure}[t]
	\centering
	\includegraphics[trim={4cm 0cm 0 10mm}, width = 16cm]{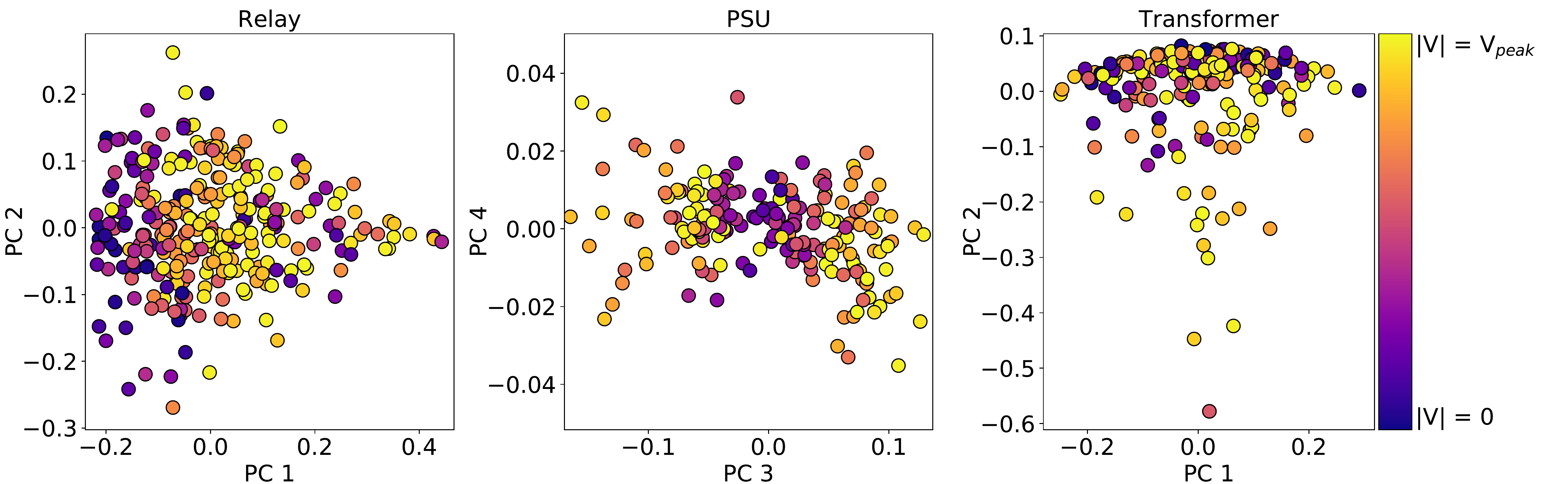}
	\caption{Kernel PCA applied separately to the RFI events from three different sources. Each marker represents a single RFI event, while the colour of the marker indicates the absolute instantaneous mains supply voltage. In the case of the power supply and the relay, there is a visible correspondence between clustering behaviour and the mains supply voltage.}
	\label{phase1}
\end{figure}

\begin{figure}[t]
	\centering
	\includegraphics[trim={2.7cm 0cm 0 25mm}, width = 16cm]{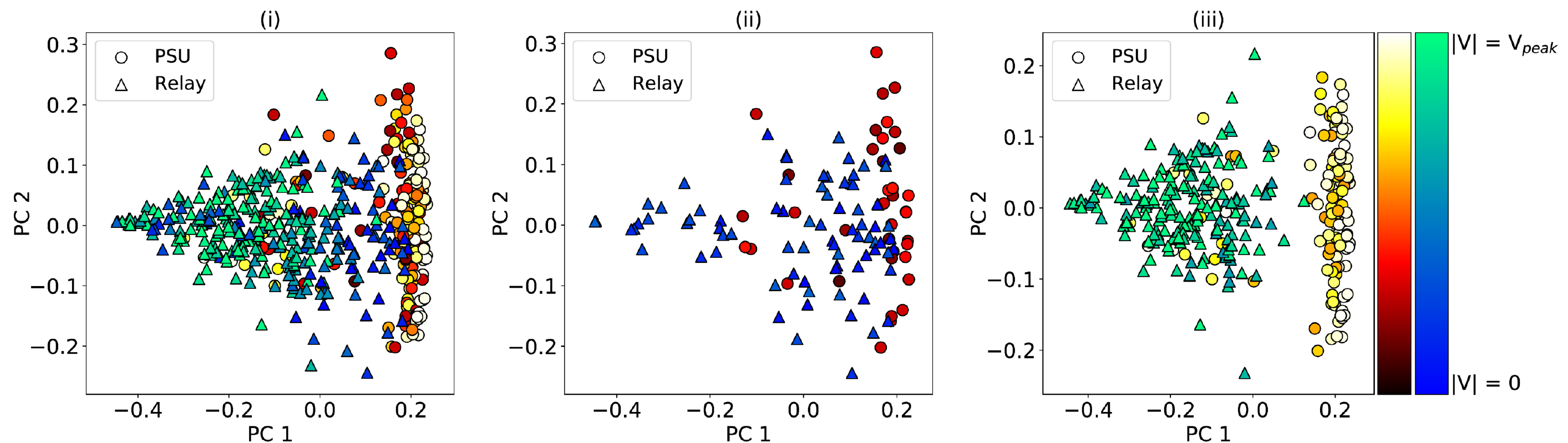}
	\caption{Kernel PCA applied to the RFI events from the relay and the power supply. Each marker represents a particular RFI event; the shape of the marker indicates its source. The colour of each marker indicates the absolute instantaneous mains supply voltage. A separate colour bar is included for each source to aid legibility. In (i), no restriction is placed on $|V|$; using the top 10 principal components, a silhouette score of 0.24 is obtained. If $|V|$ is restricted to $<0.4V_{peak}$, shown in (ii), then cluster overlap increases and the silhouette score drops to 0.07. If $|V|$ is restricted to $>0.6V_{peak}$, shown in (iii),  cluster overlap is reduced and the silhouette score increases to 0.33. }
	\label{phase2}
\end{figure}

In Fig.~\ref{phase1}, we show how the clusters of RFI transients generated by each different device change according to the instantaneous voltage of the AC supply voltage. Kernel PCA is applied to the data in a direct approach as described in Section~\ref{performance} and our previous paper \cite{s_0}. In the case of the mechanical relay and the switching PSU, there is clearly a relationship between the spread of the data points and the instantaneous supply voltage. This relationship has important implications for the classification of sources. In Fig.~\ref{phase2}, we apply kernel PCA to the data from both the PSU and the relay. Visualised in 2D, overlap between the two clusters is apparent. However, this overlap is dependent on the instantaneous supply voltage at the time of switching. We can evaluate cluster separation for different supply voltages using the mean silhouette coefficient \cite{1987sil}. The silhouette coefficient for a single example is given as follows:

$$
s = \frac{b - a}{max(a, b)}
$$

\vspace{4mm}

Here, $a$ is the average distance from the sample in question to each sample in its own cluster, and $b$ is the average distance to each point in the nearest cluster of a different class. The silhouette coefficient indicates how well-separated a point is from clusters other than its own and how well-suited it is to its own cluster. A silhouette coefficient near $-1$ indicates overlap and poorly formed clusters, while a coefficient near $1$ indicates that the points belong to appropriate, well-separated clusters.  

If only the data recorded for $|V| > 0.6V_{peak}$ are included, then the silhouette coefficient improves; this is visually apparent in the top two principal components. Conversely, if the data are restricted to those recorded when $|V| < 0.4V_{peak}$, then the silhouette score drops. Therefore, for at least some sources of transient RFI, classification accuracy is affected by the mains supply voltage at the moment of switching. By accounting for the mains supply phase, it may be possible to create better classifiers for certain sources of RFI. For example, measures of classification confidence could be increased for certain ranges of the mains supply phase.

\section{Conclusion}

We have developed a novel, effective approach to classifying and identifying impulsive RFI events according to their sources in the time domain. A useful  analogy may be drawn between these RFI events, which consist of sequences of transients, and spoken words, which consist of sequences of phonemes. We demonstrate how a dictionary of labelled transients may be constructed from which any full RFI event may be represented as a sequence of labels. In addition, we present an automated algorithm for reliably extracting individual transients from recordings of full RFI events. We show how a dictionary built in this manner may be used to identify the sources of transient RFI using hidden Markov models, in an approach loosely inspired by speech processing. This approach yields better accuracy than other well established, more direct methods when applied to an existing dataset of transient RFI. Furthermore, we show that the mains supply phase at the instant an RFI source is triggered can have a direct influence on the shape of its clusters in the principal components domain. We illustrate the implications of this effect on cluster separability, providing some insight into the difficulty of accurately identifying sources of transient RFI. In future, we expect the dictionary approach to provide a useful first step prior to the application of various different identification algorithms, in addition to hidden Markov models.

\section{Acknowledgments}

The financial assistance of the South African SKA project (SKA SA) is hereby acknowledged. Opinions expressed and conclusions arrived at are those of the author and are not necessarily to be attributed to the SKA SA (www.ska.ac.za). 

Computations were performed using facilities provided by the University of Cape Town's ICTS High Performance Computing team: http://hpc.uct.ac.za

\end{document}